\documentclass[preprint]{elsarticle}

\usepackage{hyperref}

\usepackage{tikz}
\usepackage{amsmath, amssymb, amsfonts}
\newtheorem{definition}{Definition}
\usepackage{mathtools}
\usepackage{graphicx}
\usepackage[export]{adjustbox}
\usepackage[percent]{overpic}

\usepackage{todonotes}

\usepackage{cleveref} 
\let\cref\Cref

\journal{xx}

\begin{document}

\begin{frontmatter}

\title{Mapping Research Trajectories}
\tnotetext[mytitlenote]{Fully documented templates are available in the elsarticle package on \href{http://www.ctan.org/tex-archive/macros/latex/contrib/elsarticle}{CTAN}.}

\author{Bastian Schäfermeier, Gerd Stumme, Tom Hanika}
\address{
\{schaefermeier, stumme, hanika\}@cs.uni-kassel.de \\
Knowledge and Data Engineering \\
University of Kassel \\
Wilhelmshoeher Allee 73, 34121, Kassel, Germany}

%
%

\begin{abstract}
Steadily growing amounts of information, such as annually published scientific papers, have become so large that they elude an extensive manual analysis. Hence, to maintain an overview, automated methods for the mapping and visualization of knowledge domains are necessary and important, e.g., for scientific decision makers. Of particular interest in this field is the development of research topics of different entities (e.g., scientific authors and venues) over time. However, existing approaches for their analysis are only suitable for single entity types, such as venues, and they often do not capture the research topics or the time dimension in an easily interpretable manner.

Hence, we propose a principled approach for \emph{mapping research trajectories}, which is applicable to all kinds of scientific entities that can be represented by sets of published papers. For this, we transfer ideas and principles from the geographic visualization domain, specifically trajectory maps and interactive geographic maps. Our visualizations depict the research topics of entities over time in a straightforward interpretable manner. They can be navigated by the user intuitively and restricted to specific elements of interest. The maps are derived from a corpus of research publications (i.e., titles and abstracts) through a combination of unsupervised machine learning methods.

In a practical demonstrator application, we exemplify the proposed approach on a publication corpus from machine learning. We observe that our trajectory visualizations of 30 top machine learning venues and $\sim$1000 major authors in this field are well interpretable and are consistent with background knowledge drawn from the entities' publications. Next to producing interactive, interpretable visualizations supporting different kinds of analyses, our computed trajectories are suitable for trajectory mining applications in the future.
\end{abstract}

\end{frontmatter}

\section{Introduction}
Scientometric methods provide a graspable overview over the excessive, ever-growing number of scientific publications. One particular task in this realm is the
visual \emph{mapping of knowledge domains} such as research or patent data. \citet{mappingknowledgedomains} highlighted in their foundational work the need for such maps. For their creation, a multitude of methods from different scientific areas is required. The ultimate goal here is to produce (interactive) map applications, which allow for different perspectives, selectable through user interaction. For example, a useful map for the analysis of research dynamics should reflect the changes of importance of scientific topics in time.

Automatically deriving maps, which visualize research topics over time in an interpretable, interactive manner can be a difficult task. Existing approaches are often not optimal. One particular problem to solve is creating a comprehensive, interpretable overview over the research topics present in a specific publication corpus. Many approaches for this rely on frequent terms or term co-occurrences~\cite{vosviewer, mapsofcs, 60years}. However, this limits the comprehensiveness and interpretability of the results. 
As a further limitation, previous works often visualize only one entity type, such as papers \cite{luhmann2021digital, 60years} or scientific venues \cite{Mimno2012}. In particular, author entities are seldom considered at all. Users of a scientific map could benefit if multiple entities were brought together into a single representation. This would allow a user to find thematically related authors, papers and venues by proximity in the map. Finally, the time dimension is often not addressed sufficiently. A common approach is to visualize discrete consecutive time steps~\cite{luhmann2021digital, Mimno2012}. However, a user's ability to read and comprehend the topical changes in time would be greatly aided if these steps were combined into one common map. 



In this work, we propose a novel approach for mapping \emph{research trajectories} based on topic models. This method overcomes the  depicted disadvantages and does automatically visualize different research entities, i.e., publications, authors or venues in a common temporal topic map.
Our framework for the generation of maps takes a corpus of documents for which a topical representation is computed. For this we employ \emph{non-negative matrix factorization}, which preserves human explainability due to its additivity.
Our resulting maps allow for a unified view on all entity types. Our overall goal is to obtain human-graspable and explorable visualizations. 
A particular advantage of our method is the feature to follow trajectories of research entities, such as authors, conferences or journals, in topic space over time.

We demonstrate the capabilities of our framework on a corpus of about 352,000 machine learning publications. Based on the created map, we are able to follow the trajectories of authors and venues through different subfields over time. Specifically, we map the research trajectories for several top machine learning scientists and venues, such as the \emph{NeurIPS} and \emph{WSDM} conference. All presented snapshots are taken from a running demonstration application.\footnote{\url{https://sci-rec.org/maps}}

\section{Related Work}
A PNAS journal special issue introduced the field \emph{mapping knowledge domains}~\cite{mappingknowledgedomains}, which is concerned with the automated analysis and visualization of knowledge. The contributions in that journal publication can be regarded as foundational works in this field. The collected works are comprised of articles on discovering research topics through topic models, on the analysis of scientific collaborations and of articles on visualization principles, among others. In one particular article \citet{skupin2004:theworld} visualizes research landscapes extracted from paper abstracts based on self-organizing maps and cartographic principles. However, in contrast to entities such as authors or papers, Skupin visualizes dominant terms. The editors of the special issue address the demand for interactive visualizations in the future. Specifically, they highlight that such maps should enable the selection of particular dimensions of the data, allow views from different perspectives and depict changes over time. Moreover, apart from such detailed views, the editors deem especially beneficial to observe particular entities in context of the full scientific landscape. Our work addresses all mentioned aspects.

\emph{Scientometrics} often involves the visualization of scientific entities. Maps naturally represent a particularly human-comprehensible type. In general, scientometric methods can be categorized into two main approaches:  First, content-based techniques which analyze textual data such as publication abstracts or keywords. Second, structural approaches that are based on e.g., citation or co-authorship networks~\cite{chen_citespace_2006}. Both methods are often combined~\cite{scholia}. Visualization approaches for research topics in scientometrics are often based on simple term frequencies. As an example, the widely used VOSviewer software \cite{vosviewer} uses co-occurrence frequencies of terms. Such methods are often not able to derive comprehensive topical structure in the input data. Rather, they highlight single, important terms and leave much of the interpretation work to the user.

\emph{Science overlay mapping} was introduced as a framework for creating scientific maps~\cite{overlaymapping}. In this, publication entities are represented as nodes in a \emph{base map}. Certain entities of interest can be overlaid over this map and may be visualized for different time steps. Cluster affiliations, e.g. to different research categories or subject areas, can be highlighted in color. This concept was initially applied to Web of Science ISI categories. Other approaches to base maps employ, e.g., geographic situations~\cite{overlaymapping_application}. A number of further tools and frameworks have been introduced, such as the science mapping library bibliometrix~\cite{bibliometrix}, the previously mentioned VOSviewer~\cite{vosviewer} or SciMAT~\cite{scimat}. An overview on tools used in informetrics and scientometrics is given in \cite{sciencemappingsurvey, informetricssurvey}.
Apart from these approaches, we put our focus on interactive maps which display research trajectories of authors and venues. 

Mapping \emph{trajectories} has a long tradition in geographic visualization and it has, more recently, also been used in the fields \emph{computational movement analysis} and \emph{trajectory mining}. An overview on some applications in these fields is given by ~\citet{mappingtrajectories, laube2014computational, trajectorymining}. For visualization of geographic trajectories \emph{flow maps} are often used, in particular, to depict \emph{aggregated} movement of entities~\cite{mappingtrajectories}. One very early and prominent example here is the visualization of Napoleon's Russian campaign by Charles Minard. \emph{Trajectory maps}, in contrast to flow maps, visualize the movement of single entities. Movements are depicted through line segments or arrows connecting the position fixes in a data set. 
In our work, we employ such trajectory maps to visualize scientific entities alike physical entities in geographic space. Our work translates this idea to a ``non-physical'' space in which the dimensions represent research topics. We base this on our work~\cite{schaefermeier2020topic}, in which we analyzed \emph{topic space trajectories} of scientific venues. In this work, we create a unified framework for the interactive analysis of scientific trajectories through maps. Our proposed trajectory method for scientific analyses is, to the best of our knowledge, novel.

There are further works similar to the topical maps we envision. A procedure for creating a scientific map based on term similarities, multidimensional scaling, and clustering has been introduced by \citet{mapsofcs}. This approach was applied to the field of computer science. \citet{luhmann2021digital} developed an interactive map depicting research publications. This was based on the topic modeling method \emph{LDA} \cite{lda} and the dimension reduction method \emph{UMAP} \cite{umap}. Topic stability is achieved through agglomerative clustering of topics from repeated LDA runs instead of using a more stable~\cite{topicstability} topic model method, such as non-negative matrix factorization, directly. Different from our work, research categories here are not derived automatically from the paper abstracts and the topic model. The time dimension and different entities other than papers are not considered. A further work into a similar direction is the interactive \emph{``60 years of AI research''} application~\cite{60years}, which, however, puts strong focus on the most influential papers and is limited with respect to the interactive elements we envision. None of the mentioned approaches involve trajectories.
%
%

\section{Problem Description}
For scientists, in particular scientometric analysts, it becomes more and more difficult to maintain an
overview over the development of research topics. Driving factors for this are the growing
number of publications per year and the large number of
journals, conferences and authors. Furthermore, the specialization on different fields advances over time. The sheer mass of
scientific entities, such as authors, publications and venues, hinders researchers to become aware of relevant scientific work. Moreover, different fields of research (i.e., topics) emerge over time and become more and more popular. Other fields are, more or less, replaced when new approaches are able to solve the same tasks in a more effective way. Investigating research output may often require an analysis on many different levels. These levels, e.g., may be based on different entity types, concern different research topics and regard their development over time. It may require substantial manual effort to bring all these dimensions together.

A natural consequence is the need for automated methods and tools which support researchers in these tasks. In ~\citet{schaefermeier2020topic} we developed a novel method for the examination of \emph{topic space trajectories} for the special case of scientific venues (i.e., conferences or journals). This method enables the discovery and tracking of research topics in a publication corpus based on textual contents.
From this study we conclude that the ultimate goal for the topic-based analysis of publication corpora is to create a unified view on all entities over time. This view, as argued by \citet{mappingknowledgedomains} requires to be interactive and human-graspable. This results in the \emph{problem} of automatically creating topic maps from scientific corpora.
In such maps topical similarity should be
indicated by closeness of items (i.e., entities). The user shall be able to put focus
on different aspects of interest. For this purpose, the user interface should offer at least the following functions: zoom, pan, filter and search (for entities), similar to geographical
map software. The reasoning here is that users are accustomed to such an
interface. Hence, it is natural for them to acquire knowledge in this manner, in particular, to
identify relevant information. In all this, it is essential to incorporate the temporal aspect of the corpus data.

For our \emph{solution} we especially add to these requirements human-graspability for the whole map creation process. Moreover, we require from a potential solution that the user is able to explore a corpus based on topical similarity, for example, through querying with entities, such as author, publications or venues.
Furthermore, we refrain from incorporating background-ontologies into the map-creation process. Our ratio is that this allows for the application to smaller research fields, where such an ontology is not available.
Finally and most importantly, our solution must be able to depict individual temporal entity trajectories in the common topic space.


\section{Methodology}\label{sec:method}
Our approach, as depicted in \cref{fig:mapmethod} is based on the extraction of \emph{topic space trajectories} from a corpus of publication documents. The dimensions of this space, determined through a \emph{topic model} based on a non-negative matrix factorization, indicate different \emph{research topics} where entities, such as publications, may be located. Some entities, such as authors and venues, may change their location over time, which indicates a change in their research topics. As an important aim, similar locations in this space should indicate similar research topics (or combinations thereof). 

We compute topic space trajectories for authors (i.e., scientists) and venues (i.e., conferences and journals), following the approach from \citet{schaefermeier2020topic}, which we will outline briefly here. Proceeding from this, we explain the additional steps for the computation and visualization of the final interactive map. We will substantiate the method descriptions with illustrative examples.


\usetikzlibrary{shapes,arrows}
\tikzstyle{proc} = [rectangle, draw, minimum height=2.5em, text width=6.3em, text centered]
\tikzstyle{res} = [minimum height=2em, text width=5.5em, text centered]
\tikzstyle{line} = [draw, -latex']

\begin{figure}
\begin{center}
\scalebox{0.7}{
\begin{tikzpicture}[node distance = 6.5em, auto, scale=0.10]
\node [res] (corpus) {Publication Corpus $\mathcal{D}$};
\node [proc, above of=corpus, node distance=3cm] (nmf) {Non-negative Matrix Factorization};
\node [res, right of=nmf, node distance=4cm] (topiccorpus) {Topic Corpus + Topics};
\node [proc, node distance=4cm, right of=topiccorpus] (tst) {Topic Space Trajectories};
\node [proc, below of=tst, node distance=3cm] (embedding) {2d Representation};
\node [proc, right of=embedding, node distance=3.5cm] (map) {Map\\Visualization};
\node [res, above of=map, node distance=3.5cm] (mapresult) {Interactive Map};
\path [line] (corpus) -- (nmf);
\path [line] (nmf) -- (topiccorpus);
\path [line] (topiccorpus) -- (embedding);
\path [line] (topiccorpus) -- (tst);
\path [line] (tst) -- (embedding);
\path [line] (embedding) -- (map);
\path [line] (map) -- (mapresult);
\end{tikzpicture}
}
\end{center}
\caption{Procedure for calculating topic space trajectories and visualizing them as maps.}
\label{fig:mapmethod}
\end{figure}
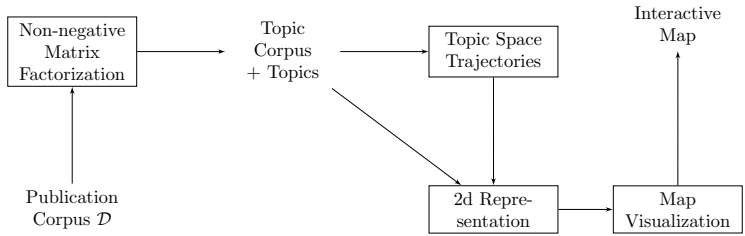

\subsubsection*{\textbf{Publication Corpus}}
The input data set for our method, as depicted in \cref{fig:mapmethod}, consists of a structured corpus $\mathcal{D}$ of publications (papers) with different information, such as author, publication venue (e.g., conference or journal) and year of publication. 

\begin{definition}[Publication Corpus]
A \emph{publication corpus} is a set $\mathcal{D} \subseteq P \times 2^A \times O \times Y $ of papers $P$ written by authors $A$, published at venues $O$ in years $Y$. A research paper $p \in P$ consists of a title and an abstract.
\end{definition}
In our example application that we introduce throughout this work, we extracted papers from the research field \emph{machine learning} from the publicly available Semantic Scholar Open Research Corpus (\emph{S2 ORC}~\cite{semanticscholar}). In more detail, we use a combination of two corpora: First, a \emph{venue corpus} of papers from the top 30 
machine learning and data mining venues listed in~\cite{kersting2019professur}. This corpus was also used in~\citet{schaefermeier2020topic}. Hence, the authors of the present work were already familiar with this data. Second, an \emph{author corpus} of papers by the top $\sim$1000 machine learning researchers. The particular authors were selected from an h-index based ranking\footnote{We used the ranking from \url{https://airankings.professor-x.de} by the university of Würzburg, Germany.} and restricted to those discoverable in S2 ORC, which had at least one publication in one of the aforementioned 30 venues. The full list of venues and authors is available in our demonstration application\footnote{\url{https://sci-rec.org/maps}}.
In total, our corpus data set consists of about 352,000 papers. In our demonstration application we visualize only a random sample of about 35,000 papers for performance reasons.

\subsubsection*{\textbf{Non-negative Matrix Factorization}}
As a first step, we extract research topics from the input corpus using a topic model based on non-negative matrix factorization (NMF). NMF is a method that finds an approximate factorization $V \approx WH$ for an input matrix $V \in \mathbb{R}_{\geq 0}^{w \times d}$, where $d \coloneqq |\mathcal{D}|$ is the number of input documents and $w$ is the vocabulary size, i.e., the number of distinct terms contained in all documents. The result matrices $W \in \mathbb{R}_{\geq 0}^{w \times t}$ and $H \in \mathbb{R}_{\geq 0}^{t \times d}$ contain representations of $t \in \mathbb{N}$ different topics as sums of word weights, as well as representations of the input documents as sums of their topics.

For the input matrix $V$, we concatenate the title and abstract of each input document. We remove stop words\footnote{Used stop word list: \url{https://github.com/RaRe-Technologies/gensim/blob/develop/gensim/parsing/preprocessing.py}} and calculate TF-IDF representations~\cite{tfidf}. The columns in the calculated matrix $W$ are interpretable as topics and the columns of $H$ are interpretable as representations of the input documents in a $t$-dimensional topic space. The number of topics $t \in \mathbb{N}$ is a hyperparameter of NMF, which we optimize with respect to a coherence measure $C_V$ \cite{topiccoherence}.
We employ our well-understood topic model from \citet{schaefermeier2020topic} and extend upon that work. This model was trained solely on the venue corpus and we use it to obtain topic representations for all papers from both the author and venue corpus. With respect to our requirement to achieve human-graspable results, we may note: An advantage of NMF for our application over other topic modeling methods, such as \emph{LDA} \cite{lda} used in \cite{luhmann2021digital}, is the stability of the computed topics~\cite{topicstability}. Hence, the reproducibility of mapping results is improved. For further advantages we refer the reader to~\citet{schaefermeier2020topic}.

In our venue corpus, NMF found 22 topics where the three top-weighted terms are, e.g., $\{\text{inference, models, bayesian}\}$, $\{\text{network, networks, neural}\}$, \linebreak$\{\text{clustering, cluster, clusters}\}$ or $\{\text{policy, reinforcement, agent}\}$. For an expert on machine learning, these four research topics are easily identifiable as \emph{Bayesian inference}, \emph{neural networks}, \emph{clustering} and \emph{reinforcement learning}. Altogether, the determined 22 research topics were labelled manually by assigning a name to it. 
For the full list of topics, refer to~\citet{schaefermeier2020topic}.

\subsubsection*{\textbf{Topic Space Trajectories}}
\begin{figure}
\includegraphics[width=0.5\columnwidth]{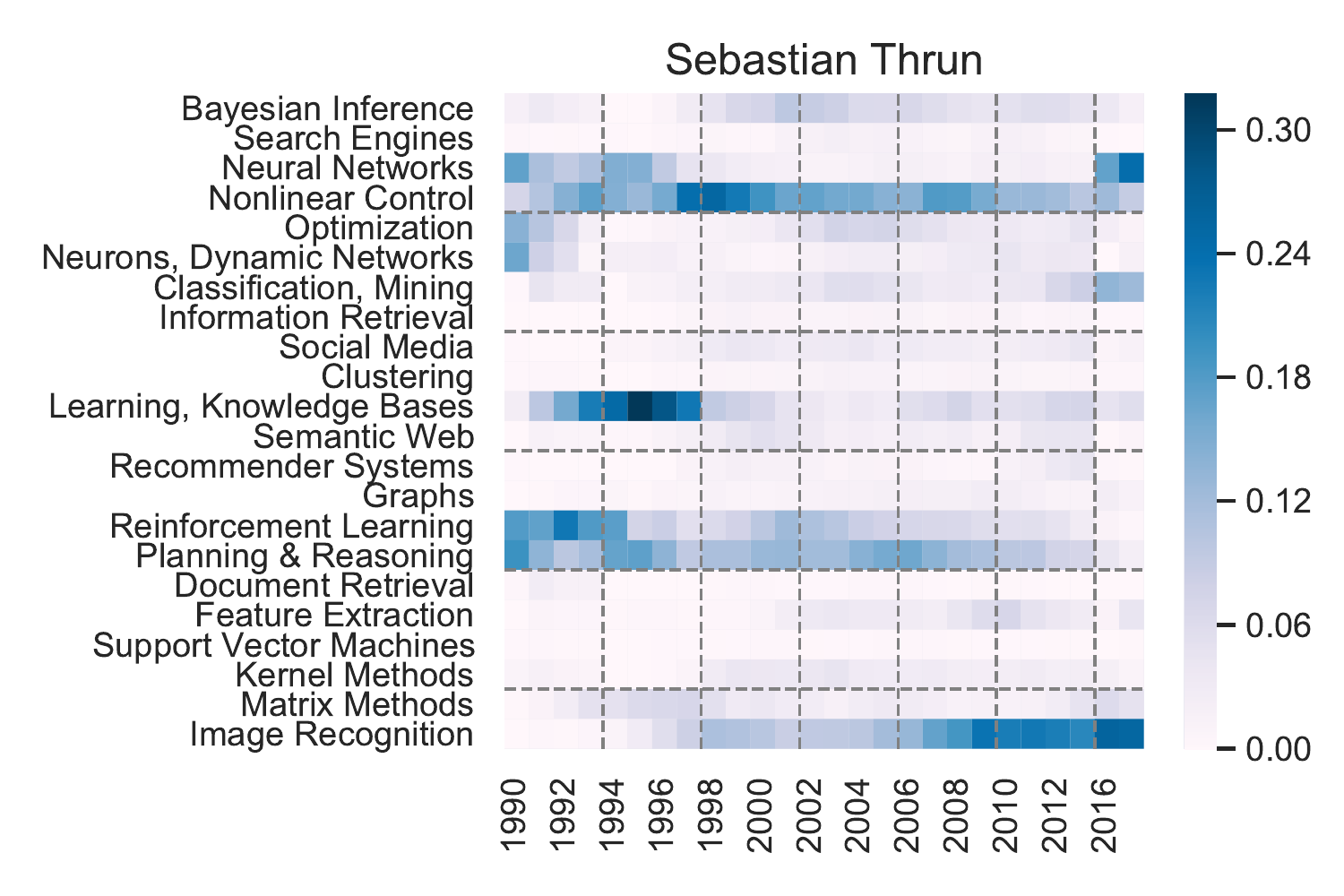}
\includegraphics[width=0.5\columnwidth]{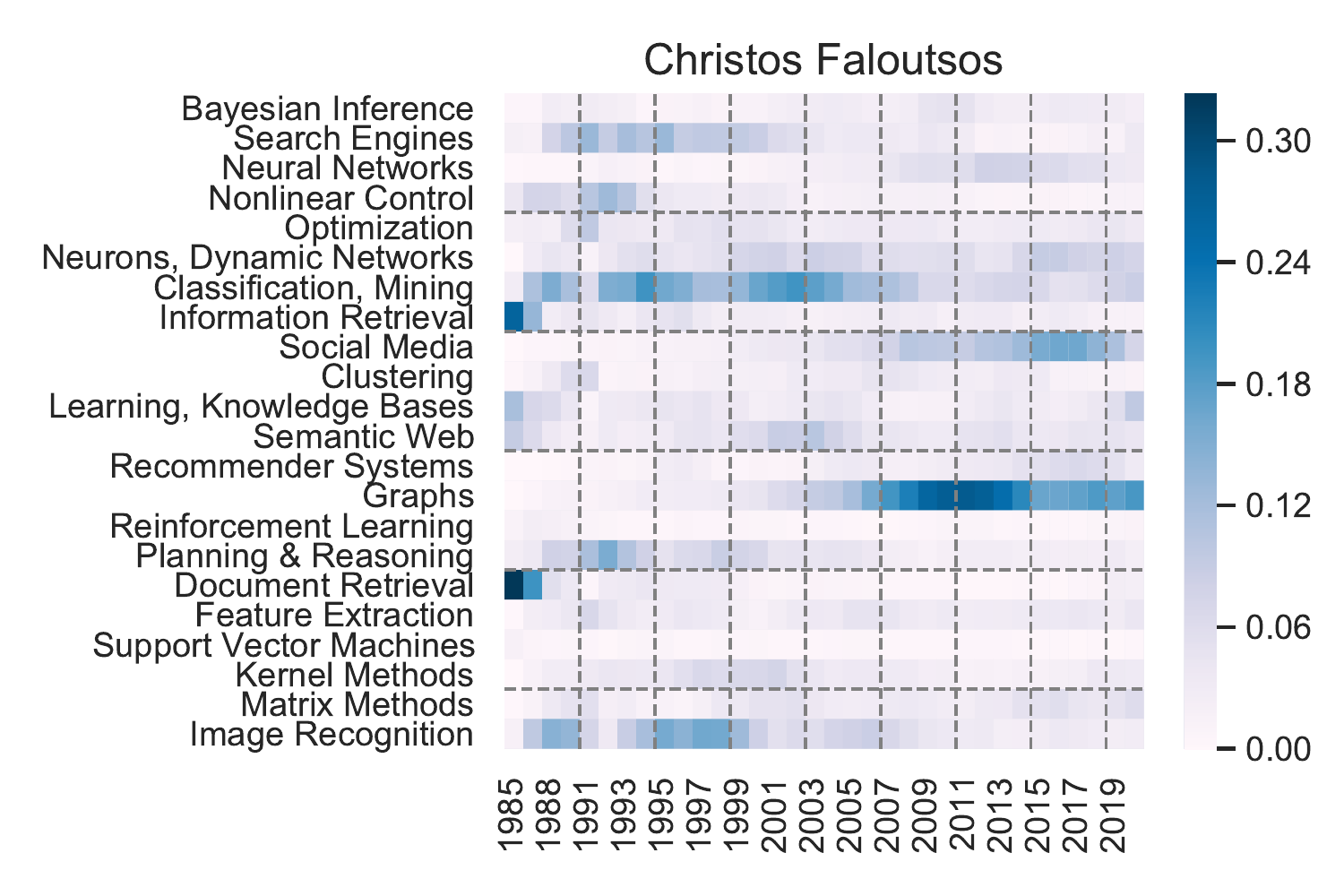}
\includegraphics[width=0.5\columnwidth]{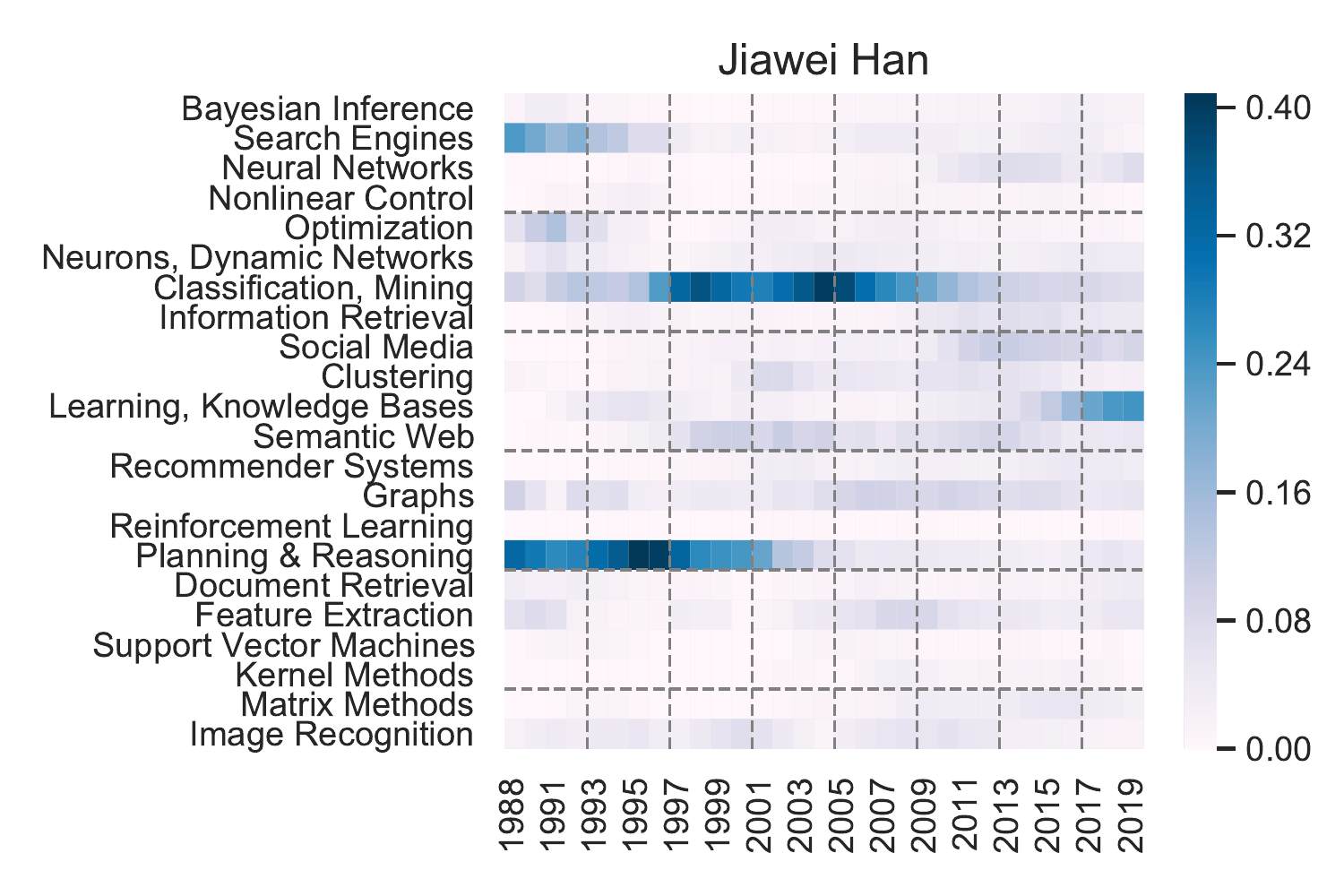}
\includegraphics[width=0.5\columnwidth]{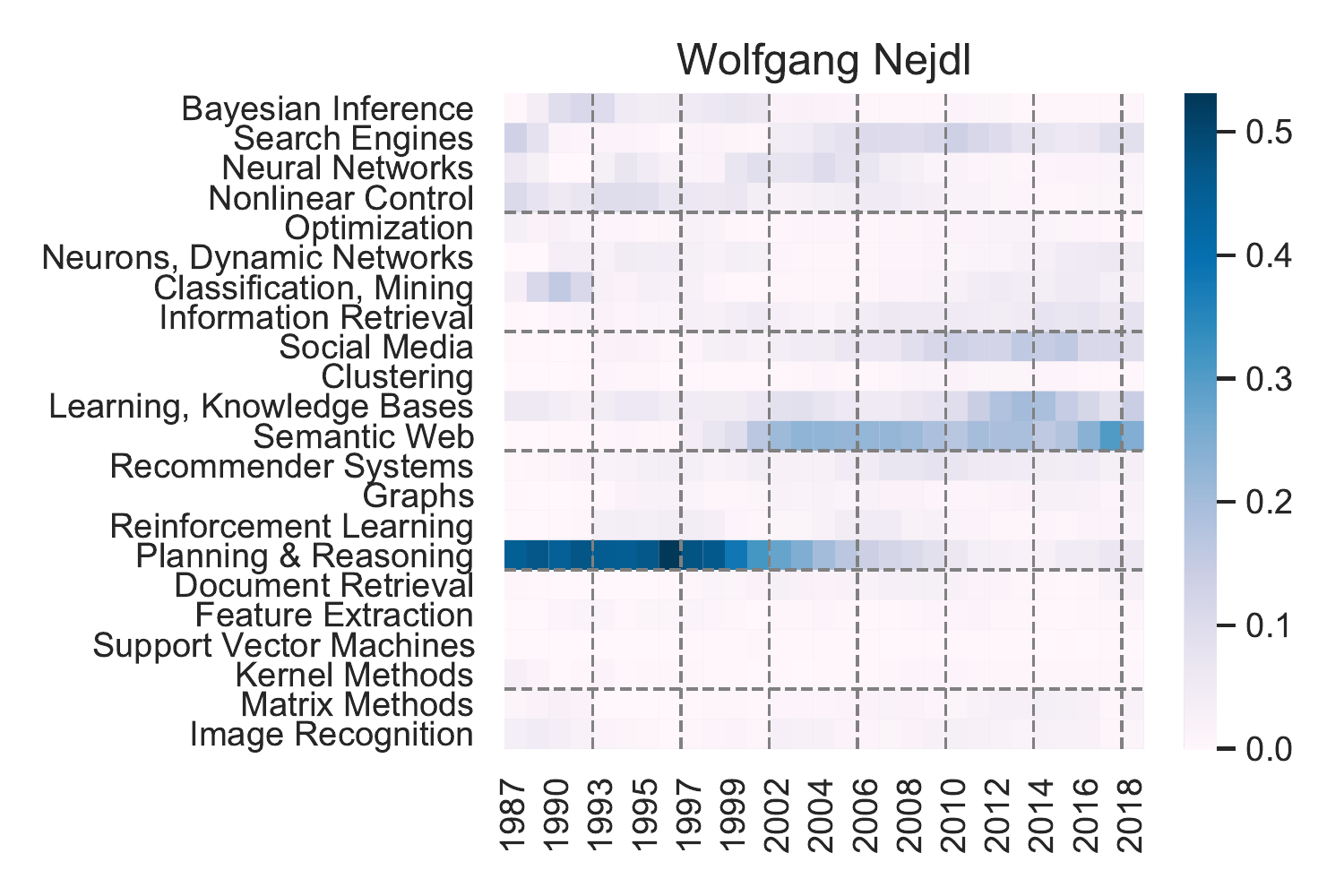}
\caption{Four examples of author topic space trajectories. The computed trajectories are visualized as heat maps. 
\emph{Note: Color scales are different per heat map since they are scaled by maximum per author. Only years with a minimum of three papers were retained.}}\label{fig:heatmaps}
\end{figure}
Based on $W$ and $H$, we obtain the topic corpus and topics depicted
in the third step of \cref{fig:mapmethod}. For each entity, i.e., \emph{author} $a \in A$ or
\emph{venue} $o \in O$ we aggregate these topical
document representations by calculating the centroids of the corresponding documents.
We calculate these centroids per publication year $y \in Y$ as well as over all years. By sorting the year-wise centroids of an entity historically, we obtain their topic space trajectories. Hence, after this step we obtain topic space representations of papers $p \in P$, of authors $a \in A$ and of venues $o \in O$. For authors and venues, we also obtain trajectories, i.e., ordered sets of topic space representations, which represent the movement of the authors or venues through topic space over time. These trajectories delineate the historical development of their focussed research topics. 
  
We reasoned that due to the common delays in the publication process authors are better represented by not only the papers written in a specific year, but also those from previous years. We therefore replace the centroid for a year $y$ by the average of the centroids from the last three years $y$, $y-1$ and $y-2$. This is equivalent to computing the moving average of a time series and leads to smoother trajectories in practice. Although there are computationally more demanding methods \cite{trajectorysmoothing}, in our experiments we find that the previously explained approach, that is also used in physical trajectory smoothing, is sufficient. For venues we only use publications from a specific year to represent the current research. In practice, venue trajectories were already smoother due to the larger number of publications.
As an example for the computed trajectories, \Cref{fig:heatmaps} depicts the results for four different authors. The computed topic weights are visualized as heat maps, in which rows indicate topics and columns indicate years. One may notice the topics in focus often stay consistent over some time and do also change smoothly. Moreover, some authors are more focussed and others are concerned with many different topics.

\subsubsection*{\textbf{Computing 2d Representations}}
\begin{figure}
\includegraphics[width=0.99\columnwidth,trim={1.2cm 0cm 3cm 1.5cm},clip]{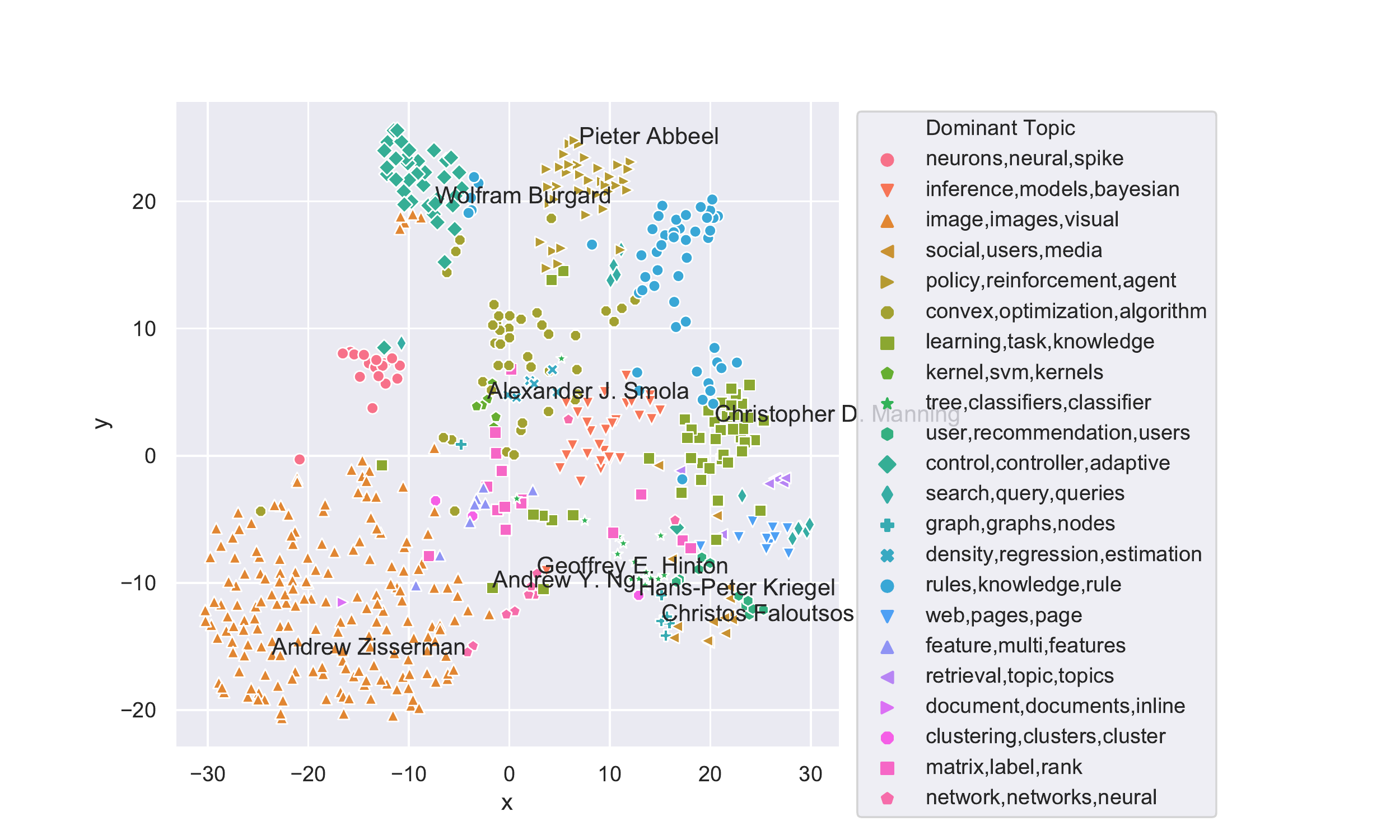}
\caption{Two-dimensional representations of authors. Each marker represents an author. Marker colors together with shapes indicate the respective main topics. For many topics, clusters emerge. Coordinates were computed through t-SNE. The locations for some well-known authors from different fields are highlighted for reference. Best viewed in color.}\label{fig:authormap}
\end{figure}
In the penultimate step, we apply a dimensionality reduction method to derive
a $2$-dimensional representation
from our $t$-dimensional topic vectors.
We use the coordinates of this $2$-dimensional representation for the final
visualization in a map. While more detailed topical information is
lost here, such visualizations are much easier to interpret, since
closeness in a 2-dimensional coordinate system can be interpreted as
topical similarity. In \citet{schaefermeier2020topic},
multidimensional scaling (MDS) was studied for dimensionality
reduction of similar data. However, in practice MDS is infeasible for
large input corpora due to the quadratic complexity for recalculating the pairwise distances~\cite{mdscomplexity}. Hence, in the present work, we use \emph{$t$-distributed stochastic neighborhood
embedding} (t-SNE) \cite{tsne_jmlr}, which can be computed with $O(N \log N)$ complexity~\cite{tsneaccelerating}, where $N$ is, in our case, the number of input documents. This embedding method is widely used for the visualization of high-dimensional and very
large data sets (e.g., with millions of objects by~\citet{tsneaccelerating}).

We may note that NMF is capable of computing 2-dimensional representations as well. However, this approach would limit potential representations to only two \emph{topics}. Moreover, such a reduction is limited by the linearity of NMF~\cite{nonlinearnmf}. In contrast, the nonlinear t-SNE, is capable of finding meaningful 2d visualizations for data sets that are intrinsically more complex. Hence, our approach combines the strengths of both dimension reduction methods: Deriving human interpretable topics through NMF and finding meaningful map coordinates through t-SNE for the visualization.

\Cref{fig:authormap} depicts an example of the thus calculated coordinates for the $\sim$1000 machine learning authors. These representations were computed in preliminary experiments over all publication years. In addition to displaying coordinates, we highlight in \cref{fig:authormap} each author's overall main topic (i.e., the topic with maximum weight). Noticeable, clusters emerge based on said main topics. For some well-known authors we also added labels to the right of the plot markers, for reference. Based on this, we describe our mapping process below.

\subsubsection*{\textbf{Map Visualization}}
\begin{figure}
\includegraphics[width=0.99\columnwidth]{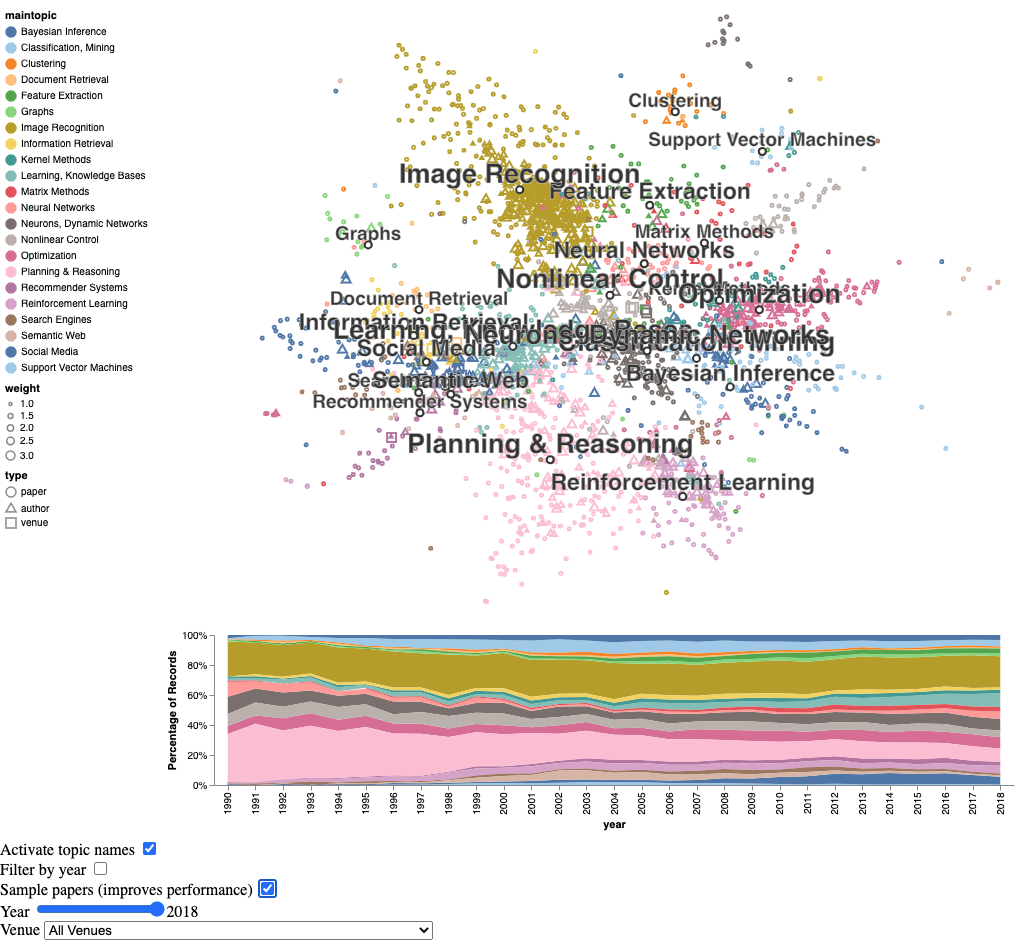}
\caption{Depiction of the scientific map approach. Entities, i.e., papers, authors and venues are displayed using different markers. Colors indicate main research topics. The stream graph at the bottom displays research prevalence of research topics over time. The legend to the left as well as the form elements at the bottom belong to the interactive interface.}\label{fig:mapoverview}
\end{figure}

For the visualization of our scientific maps, we mark each entity location in a $2$-dimensional coordinate system, as depicted in~\cref{fig:mapoverview}. We use different markers, such as dots or triangles, for different entity types. Furthermore, we color them by their main topic, i.e., the topic corresponding to the maximum weight within an entity's topic vector. To provide some points of orientation in the map we display topic names at each topic's center. These are computed as the centroids over all papers with that main topic. Different sizes of topic names indicate the number of papers in the respective fields.

In order to enable the user to explore the just illustrated map, our approach provides several interactive elements. We allow for zooming and panning within the map, similar to geographic (software based) maps. When the user's cursor hovers over an entity, such as a paper or author, further details are given in a tooltip, for example paper title, author name, main topic or the publication venue and year. 
We allow for different filtering options with respect to the entities. More specifically, the user can restrict the visualized entities to specific main topics by selecting them from the plot legend. From said legend, the user can select specific entity types to visualize or remove from the map. By means of these selections, one is able to, for example, restrict an analysis to only authors or venues. The reasoning here is that, e.g., papers may be irrelevant to some applications. The user can also restrict the visualized entities to those from a specific year. In order to follow the development of selected entities over time, we added a slider which allows to step through the publication years.

Below the map, we additionally display a stream graph that depicts the percentages of research on the different topics. Percentages are calculated based on the main topics of the data entities. Hovering over one of the colored areas from the stream graph displays the associated topic in a tooltip.

\begin{figure}
\includegraphics[width=0.99\columnwidth]{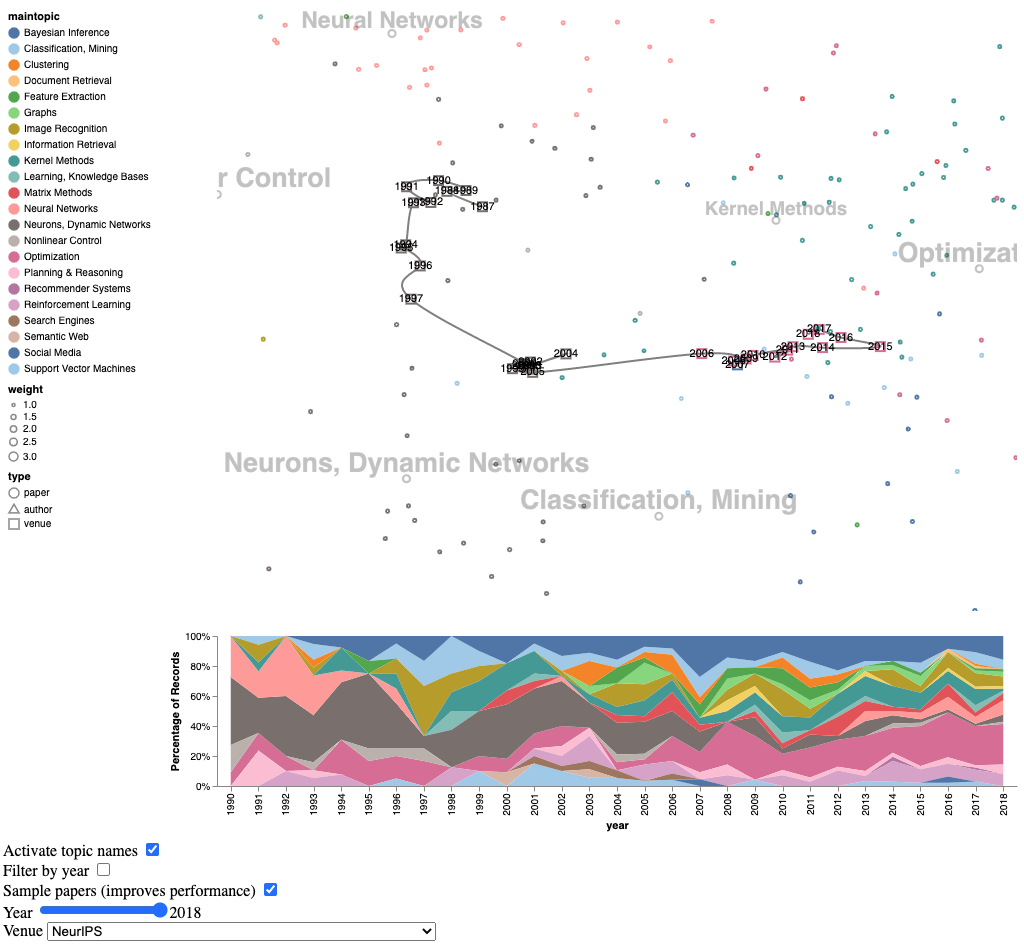}
\caption{Depiction of the trajectory of a specific entity. In this case, the \emph{NeurIPS} conference is selected. Papers published at the conference are displayed in the map additionally to the trajectory. Instead of for the full data set, the stream graph at the bottom now displays the topic distribution for only the selected entity.}\label{fig:trajectoryview}
\end{figure}

From a dropdown menu or using a search field, the user can restrict the view to a specific entity in order to analyze its trajectory.
We show the corresponding papers (written by the author or published at the venue) and depict the precomputed trajectory elements (see \cref{fig:trajectoryview}). The points of the trajectory are connected by line segments using spline interpolation. We label each trajectory point by the associated year. In the trajectory view, the stream graph displays the topics of the selected entity only instead of for the full data set.

In the practical implementation\footnote{\url{https://sci-rec.org/maps}} of our scientific map approach, we enhanced the rendering performance by displaying only a sample of the complete set of papers. The user can downsize this sample even further using a checkbox. This sampling is automatically deactivated once a specific venue or author is selected for detailed inspection. We used for our implementation the \emph{Altair} library~\cite{VanderPlas2018}.

\section{Trajectory Maps: A Case Study on Machine Learning Entities}
In our demonstrator implementation, we exemplify the explained trajectory mapping approach on a corpus of $\sim$352,000 machine learning publications with the trajectories of $\sim$1000 authors and 30 venues (for further details, see \emph{Publication Corpus} in \cref{sec:method}). In the following, we analyze different example trajectories and point out notable observations that can be made in the computed maps. For this, we will first give an overview on the layout of topics. 

\subsection*{Analysis of the Entire Map}
In the overall map layout we obtain, the entities form clusters regarding their main topics, as visible in \cref{fig:mapoverview} (and as we already observed in our preliminary experiments in \cref{fig:authormap}). Hence, different areas in the resulting map can, to some degree, be assigned to different research areas.
In the layout of the topics and their labels, we observed that semantically related research areas are often located closely nearby to each other. This is an expected result, given that the objective of the dimension reduction method is to preserve the vicinity from the original topic space. As a consequence, more similar, and therefore possibly correlated, topics are laid out more closely together. As a further implication resulting from this, more general topics are located centrally between related, but more specialized topics. For example, in the left area of the map in \cref{fig:mapoverview} we find the \emph{Information Retrieval} topic, which is surrounded by topics such as \emph{Document Retrieval}, \emph{Social Media}, \emph{Semantic Web} and  \emph{Recommender Systems}. Furthermore, the center of the overall map contains the topic \emph{Classification Mining}, which represents the highly general \emph{data mining} field that is strongly related to all other topics.

We may note that some topics researched by authors in our data set are not explicitly captured by our employed topic model instance, since it was only trained on the machine learning venue corpus. Examples for such topics are the fields \emph{natural language processing} or \emph{robotics}. Since our corpus does not contain publications from conferences or journals focusing on these more specialized fields, no topic was created for them. Yet, the resulting topic representations of publications in fields not captured are often located, as best as possible, in related topic areas. For example, we observed that papers from robotics are often located in the areas \emph{Nonlinear Control} or \emph{Image Recognition}. Both of these fields are often applied to robotics.

\subsection*{Individual Author and Venue Trajectories}
\begin{figure}
\begin{overpic}[fbox,width=0.5\columnwidth,trim={6cm 20.5cm 5cm 2cm},clip]{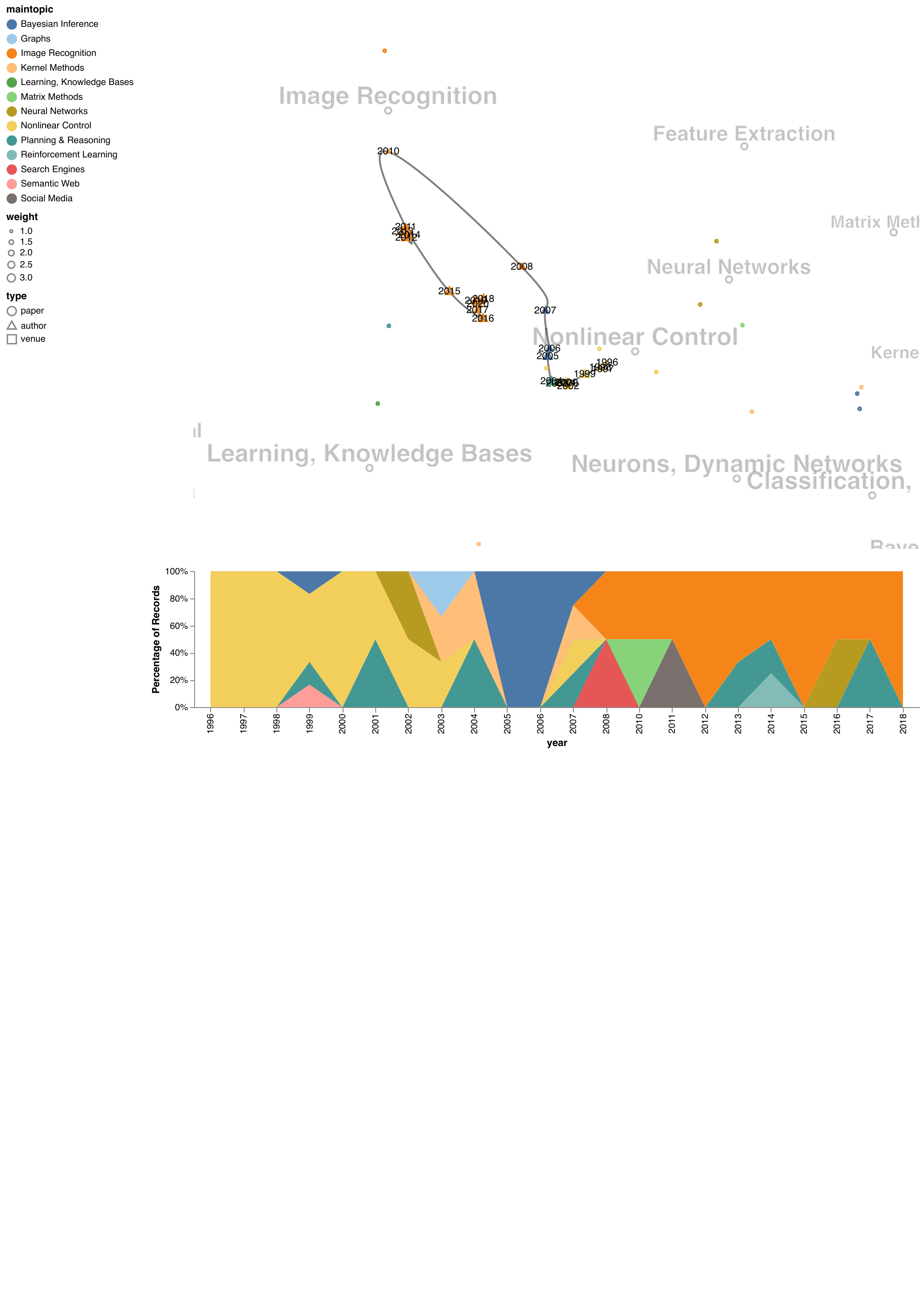}\put (5,5) {Dieter Fox}\end{overpic}
\begin{overpic}[fbox,width=0.5\columnwidth,trim={8.5cm 19cm 2cm 3.15cm},clip]{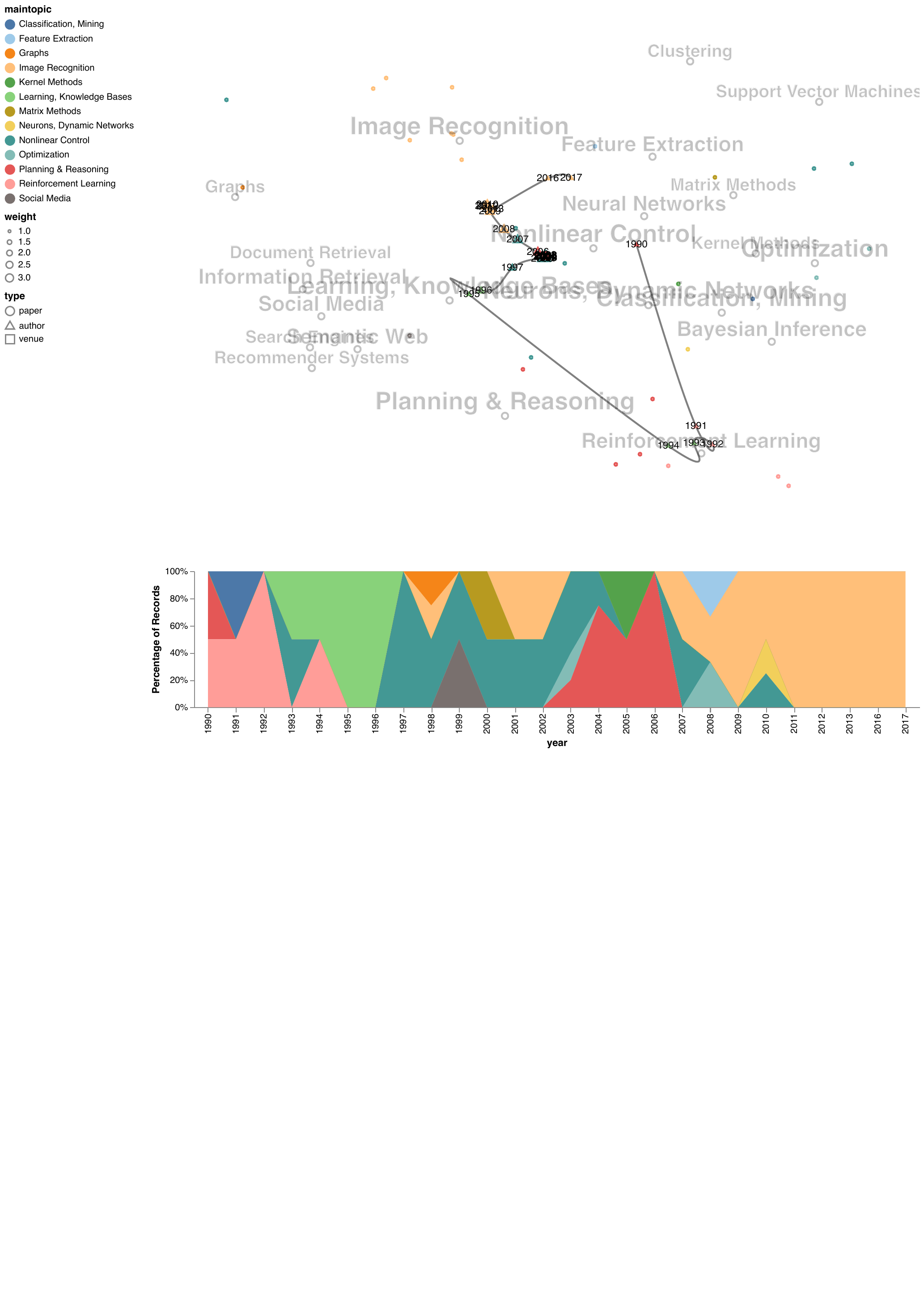}\put (5,5) {Sebastian Thrun}\end{overpic}
\begin{overpic}[fbox,width=0.5\columnwidth,trim={6cm 2.2cm 0 1.2cm},clip]{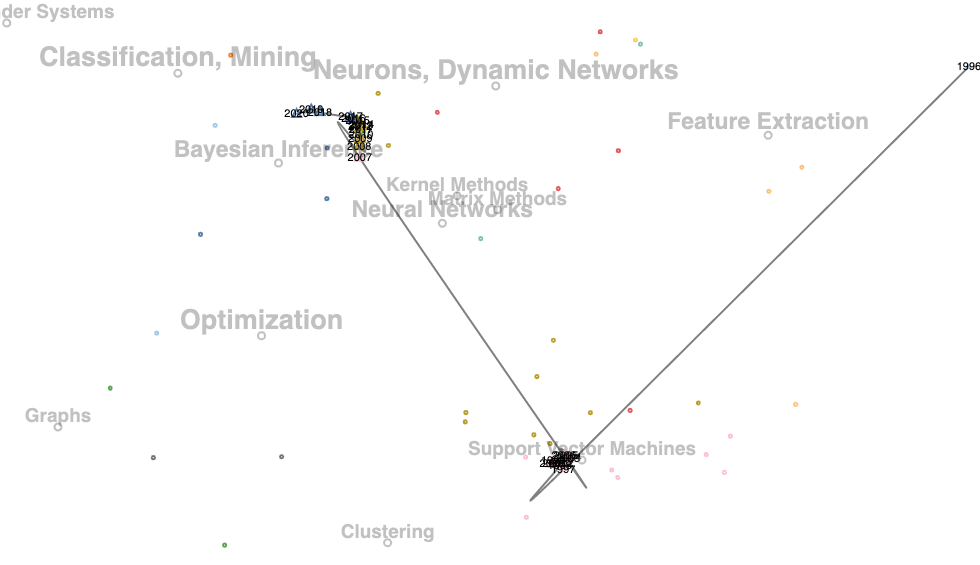}\put (5,5) {Bernhard Schoelkopf}\end{overpic}
\begin{overpic}[fbox,width=0.5\columnwidth]{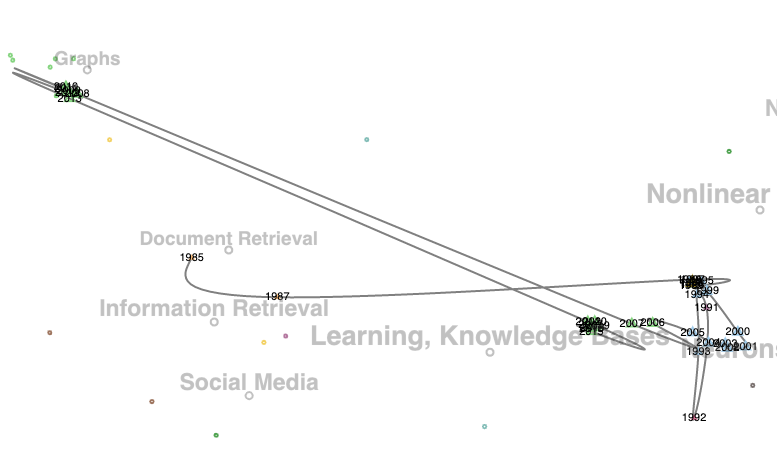}\put (5,5) {Christos Faloutsos}\end{overpic}
\begin{overpic}[fbox,width=0.5\columnwidth]{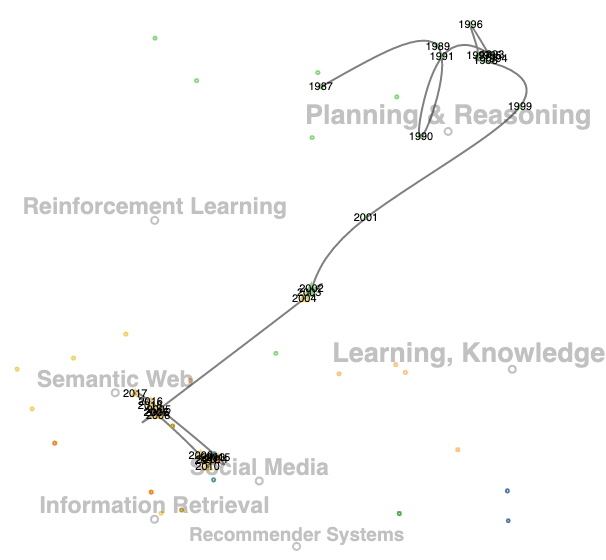}\put (5,5) {Wolfgang Nejdl}\end{overpic}
\begin{overpic}[fbox,width=0.5\columnwidth,trim={0.8cm 0 1.1cm 0.1cm},clip]{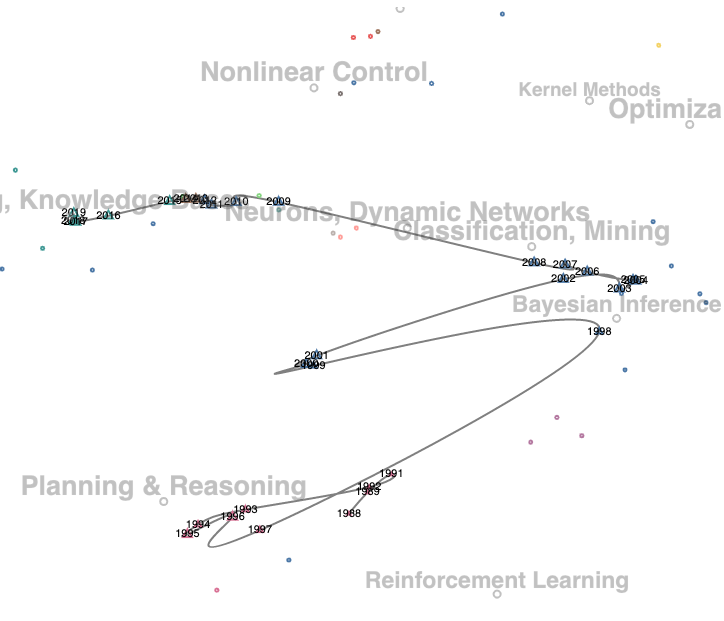}\put (5,5) {Jiawei Han}\end{overpic} 

\caption{Trajectories of six AI authors. \textbf{Top row:} \emph{Dieter Fox} (left) and \emph{Sebastian Thrun} (right). Both authors are from the robotics area and published in this field together as co-authors. \textbf{Middle Row:} \emph{Bernhard Schölkopf} (left) and \emph{Christos Faloutsos} (right). The topical focusses change from support vector machines to inference methods (left) and from general data mining to graphs (right). \textbf{Bottom row:} \emph{Wolfgang Nejdl} (left) and \emph{Jiawei Han} (right). Here the focus changes from planning \& reasoning to semantic web and social media (left) and from planning and reasoning to classification \& mining to knowledge bases (right). 
}\label{fig:authortrajectories}
\end{figure}

Our trajectory mapping approach allows for the analysis and comparison of research trajectories by different authors and venues. As an example, in~\cref{fig:authortrajectories} we present the topic space trajectories of six different machine learning authors. For a better overview, the map view in each of these pictures was zoomed and panned to the relevant area of the topic space map. Note that further details, such as main topics for trajectory points, can be inferred from the full map view (examples given in the appendix, ~\cref{fig:jiaweihancloseup,fig:michaeljordancloseup}) and, in particular, in the online demo application. Four of the depicted author trajectories were also visualized as heat maps in \cref{fig:heatmaps}.

The authors in the top row of \cref{fig:authortrajectories} are Dieter Fox (left) and Sebastian Thrun (right), which are both researchers in the robotics field. They published together, among further works, a standard work on probabilistic robotics~\cite{roboticsbook}. We may recapitulate that papers and authors in the robotics field are often located in the \emph{Nonlinear Control} area of our topic space. This topic is related to control theory in general and a reasonable element of robotics. We also observe that Dieter Fox at the later parts of his trajectory, published work in the \emph{Image Recognition} area. The titles of his publications displayed in this map area (which can be inspected through the tooltips in the interactive map) confirm this. Image recognition (i.e., computer vision) is commonly used in robotics, e.g., for visual navigation. The last part of the trajectory by Dieter Fox revisits the initial nonlinear control area. In contrast to these topics, an early part of the trajectory of his coauthor Sebastian Thrun (right) started in the reinforcement learning field, on which he published various works.

In the middle row of \cref{fig:authortrajectories} we see that the trajectory of Bernhard Schölkopf (left) in an early part of his career runs from support vector machines to Bayesian inference. By inspection of his publications in the map we found that he is conducting research on \emph{causal inference}, which our model locates in the \emph{Bayesian inference} area.
Christos Faloutsos (right) at the beginning of his career, published on document retrieval, and then extended his scope to data mining in general (which is visible in the full map view). Later he focussed on graphs and, in particular, graph mining. In the bottom row, the trajectory of Wolfgang Nejdl (left) shows that he started in the \emph{Planning \& Reasoning} area and later switched to \emph{Semantic Web} and \emph{Social Media}. Jiawei Han's research trajectory (right) started in \emph{Planning \& Reasoning}, proceeding with \emph{Classification, Mining}  and, finally, \emph{Knowledge Bases}. Again, the publications of the authors in the specific years confirm the trajectory courses.

\begin{figure}[th]
\begin{overpic}[width=0.5\columnwidth,trim={8cm 10.5cm 1.5cm 4cm},clip,fbox]{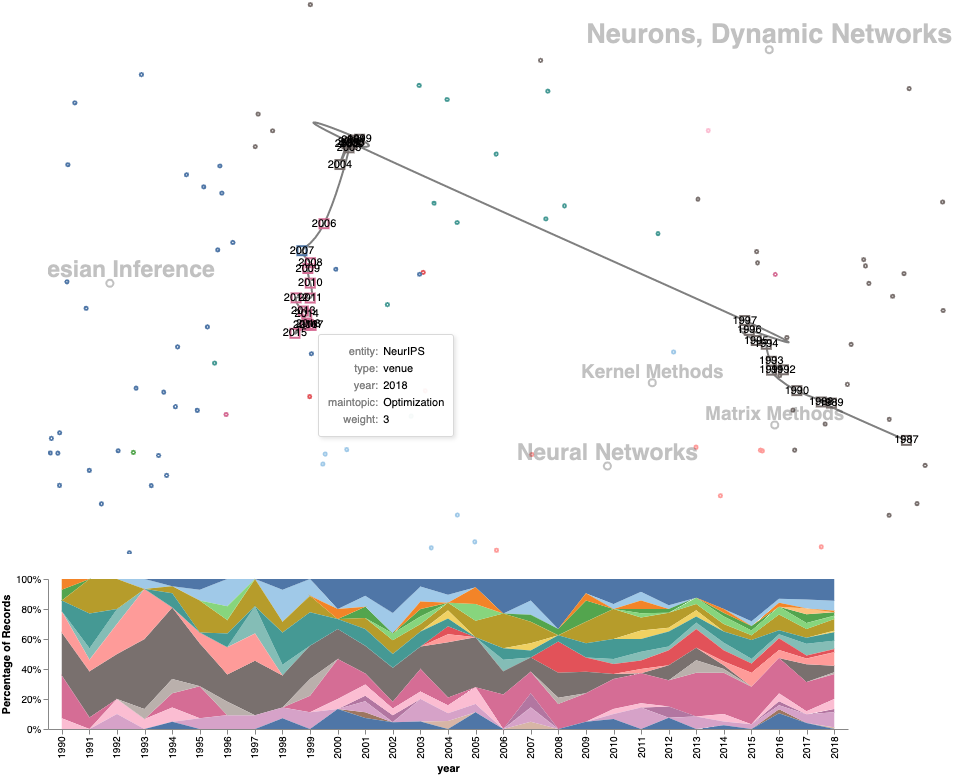}\put (5,5) {NeurIPS}\end{overpic}
\begin{overpic}[width=0.5\columnwidth, trim={8cm 9cm 4cm 7cm},clip,fbox]{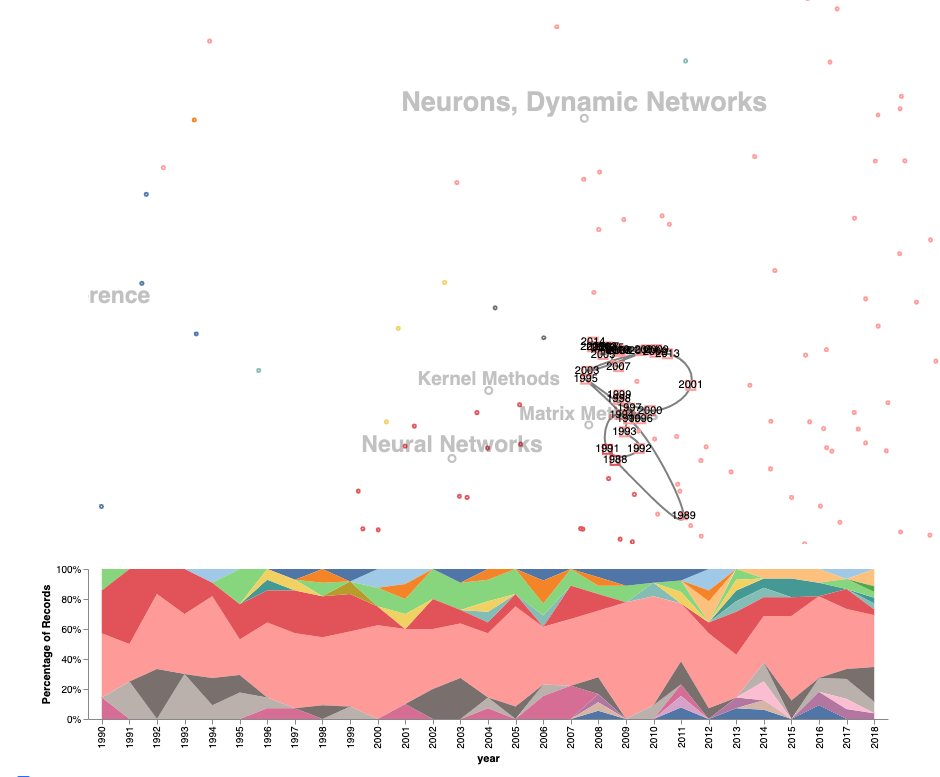}\put (5,5) {Neural Networks}\end{overpic}
\begin{overpic}[width=0.5\columnwidth, trim={5.2cm 9.5cm 1.5cm 4cm},clip,fbox]{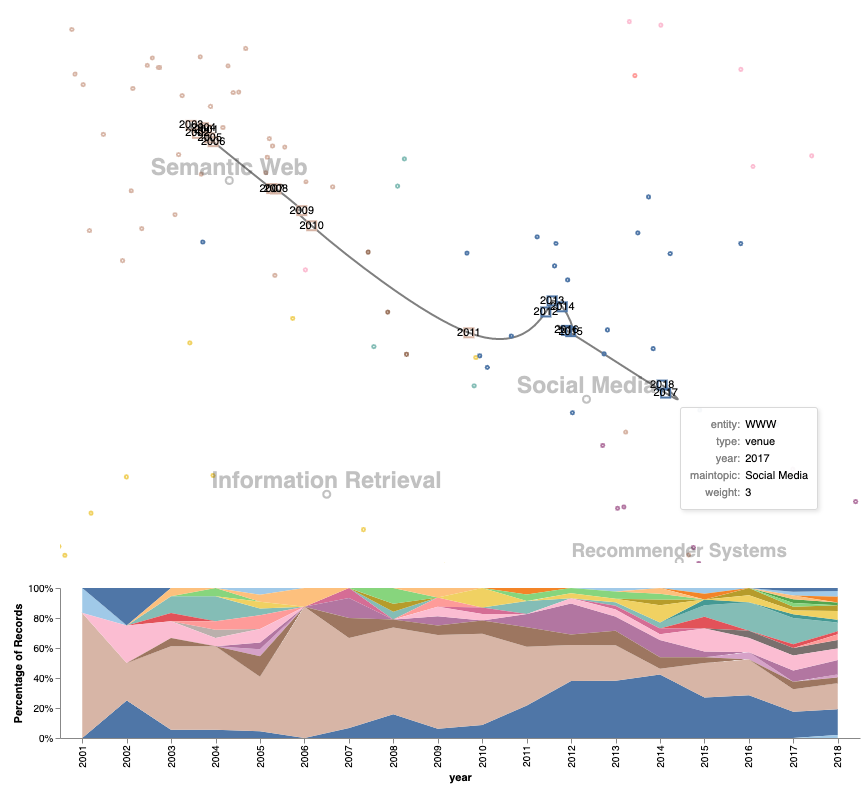}\put (5,5) {WWW}\end{overpic}
\begin{overpic}[width=0.5\columnwidth, trim={3cm 8cm 0 1.75cm},clip,fbox]{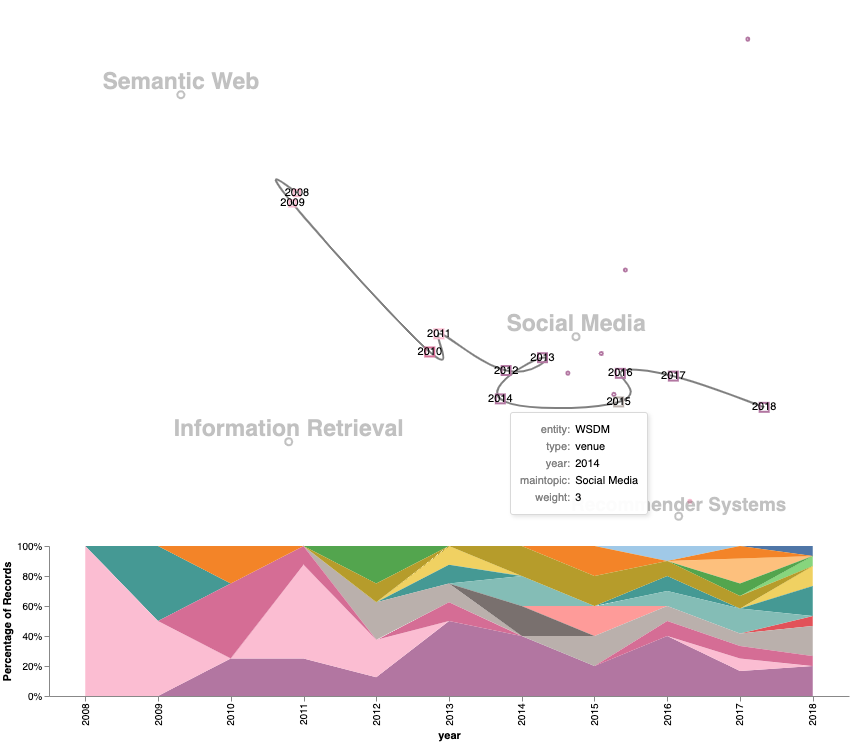}\put (5,5) {WSDM}\end{overpic}

\caption{Trajectories of different venues. \textbf{Top row:} Two neural networks venues \emph{NeurIPS} (left)
and the journal \emph{Neural Networks} (right).  NeurIPS in recent years became a more general machine learning venue with
  some focus on optimization, but also, e.g., Bayesian
  inference. Neural networks staid focussed on neural networks.
  \textbf{Bottom row:} The conferences \emph{WWW} (left) and \emph{WSDM} (right). Both moved from semantic web to social media, the latter of which became a more popular topic in general.
}\label{fig:venuetrajectories}
\end{figure}

Analogously to authors, we may analyze and compare trajectories of publication venues within the \emph{same map}. An example for four conferences or journals is depicted in~\cref{fig:venuetrajectories}. In the top row, we depict the trajectories of two venues on neural networks: The \emph{NeurIPS} conference (formerly \emph{NIPS}, left) started in the same research areas as the journal \emph{Neural Networks} but, in contrast, later broadened its focus to \emph{Optimization} and more general machine learning topics. The same pattern can be observed in ~\citep{schaefermeier2020topic}. In the bottom row, the conferences \emph{WWW} (left) and WSDM (right) exhibit a similar course from \emph{Semantic Web} to \emph{Social Media}. The latter topic became more popular overall in recent years, e.g., in social network analysis applied to platforms such as Twitter.

\subsection*{Observed Patterns}
We observed some reoccurring patterns in author trajectories that may lead to new interesting research questions. One of these patterns is that a relevant share of authors (i.e., researchers within the investigated realm) change their research focus in the early years of their career. We first attributed this to some artifact of our projection method. Although it is anecdotal, as researchers, we ourselves often experienced similar behavior with colleagues in their career. Therefore we looked closer into the publication work of the authors concerned (e.g., Christos Faloutsos) and are able to substantiate our claim for the observed trajectory pattern. Hence, we may assume that researchers are initially in an orientation phase, in which they still have to find their suitable research area. We leave open for future quantitative analysis (i.e., considering all trajectories) to estimate the frequency of the observed phenomenon.

As another observation, we were able to identify different types of research careers: In some instances, scientists stay focussed on a specific research area throughout their whole career. 
Notably, researchers from the \emph{Image Recognition} field rarely move into completely unrelated areas of machine learning, according to our observations (e.g., Andrew Zisserman and Jitendra Malik). In contrast, there are researchers with strong changes throughout their career. Often these authors follow the general research trends (i.e., hot topics), as we observed. From the late 1990s to early 2000s, for example, support vector machines became a popular hot topic, and later lost in popularity~\cite{schaefermeier2020topic}. Following these trends may be motivated by personal interests as well as incentives by research funds. The vanishing interest in certain research topics may also occur naturally due to the development of newer, more successful methodologies.

In summary, we found that the computed research trajectories provided numerous insights into the temporal development of scientists. Moreover, the results naturally lead to interesting research questions. The reoccurring patterns we observed could consecutively be investigated through the application of trajectory mining methods. As an example, trajectory clustering and stay point detection may reveal such patterns and, possibly confirm the ones we observed~\cite{trajectorymining}. Another approach that may be used to test trajectory hypotheses is the HypTrails method~\cite{hyptrails}.

\subsection*{Discussion}
Altogether, we find that our approach of mapping research trajectories leads to well-interpretable visualizations of author's research careers. In the same manner as for authors, it allows to follow the focus changes of conferences and journals over time within the same maps. A particular advantage of the trajectory approach is that the topical development over time becomes apparent through one single visualization that can be interpreted rather quickly.

Our practical map implementation with tooltips and visualizing papers close to a segment of a trajectory give additional detail information to a user. In particular for these elements the interactive map application is a helpful solution for supporting users in their analysis tasks. The interactive map allows to pan and zoom to different segments of a trajectory. This facilitates obtaining an overview on an entity in context of the overall topic space as well as in context of details in the closer topic neighborhood, an important overall goal in the mapping of knowledge domains brought up by \citet{mappingknowledgedomains}.

As a limitation, the representation of entities through average and maximal topic weights may occasionally oversimplify their overall topical distribution. This may become a problem when authors or venues have a very broad distribution of focussed topics at a point in time. The corresponding trajectory points may then be positioned between different topic areas in the map and are thus more difficult to interpret. In these cases, the tooltips, which give information on the main topic, as well as the stream graph below the map may give helpful further information. Often, however, we also observed that such broader topic distributions were positioned at the center of various topics related to each other or in a very general topic area such as \emph{Classification, Mining}, which can be a reasonable description for research related to many different topics in the field.

\section{Conclusion and Outlook}
In this work, we introduced a principled approach for mapping research trajectories of authors and venues. Our ideas draw from methods from conventional ``physical'' trajectory analysis and geographical information visualization. We demonstrated the practical applicability of our proposed method in an interactive map application that can be experimented with by the reader.\footnote{\url{https://sci-rec.org/maps}} We analyzed examples of entity trajectories and found them to be consistent with background knowledge from author and venue publications.

We think that our approach opens new possibilities for mapping and analyzing research trajectories for  various other research entities. For example, our idea is applicable to specific research institutions, groups and communities, e.g., based on universities, countries or author conglomerations. Besides research, maps could also be computed for completely different, document-based knowledge domains, such as social media posts and their authors, news articles or patent data. As a further possibility, techniques from trajectory mining are applicable to our computed trajectories. The re-occurring patterns we observed in author trajectories allow for the generation of new hypotheses that could be investigated further through trajectory clustering, stay point detection, sequential pattern mining or HypTrails~\cite{trajectorymining, hyptrails}.

We also identified some limitations of our approach. The dimension reduction to two dimensions can occasionally lead to  artifacts, i.e., some points being projected to a location distant from the topic cluster where we would expect it. 
For our trajectories, more stability in this regard might be achieved by averaging the 2d coordinates computed for several random perturbations of the original topic vector. Another idea here would be to compute centroids based on the computed 2d coordinates of documents (instead of, as we did, computing centroids of documents followed by the reduction to 2d). In this case, some adjustment of centroid computations might be required due to the non-linear projection method. For the practical applicability in productive settings the performance of the map computation as well as the clarity of the visualization should be improved. For this, we experimented with a level-of-detail based visualization, where only highly relevant entities are depicted in the full map overview and details added  gradually once the user zooms closer into the map. This will enable the user to maintain the ``big picture'' as well as to steadily obtain a more fine-grained view on specific topic areas. We plan to base the levels of detail on relevance weights of entities, which we compute from paper numbers and additional citation data. Finally, the used topic model does not capture all research topics in our data set, since it was trained only on a subset of the documents (i.e., the venue corpus as explained in \cref{sec:method}). With this intentional decision we wanted to achieve comparability of our results to an earlier work~\cite{schaefermeier2020topic}. However, in productive applications, the topic model should be computed from the entire data set that is visualized.

For future research we envision the usage of a hierarchical topic model that captures topics and subtopics~\cite{hierarchicaltopicmodel}. These subtopics would complement our aforementioned level-of-detail suggestion and could be gradually refined in the visualization upon zooming. Moreover, we suggest to improve the automation of the map building process through an automated topic labeling procedure, e.g., through frequent n-grams or based on the FREX score~\cite{frexscore}.



\bibliography{paper}

\newpage
\appendix
\section{Appendix}
\begin{figure}[h]
\includegraphics[width=0.99\columnwidth]{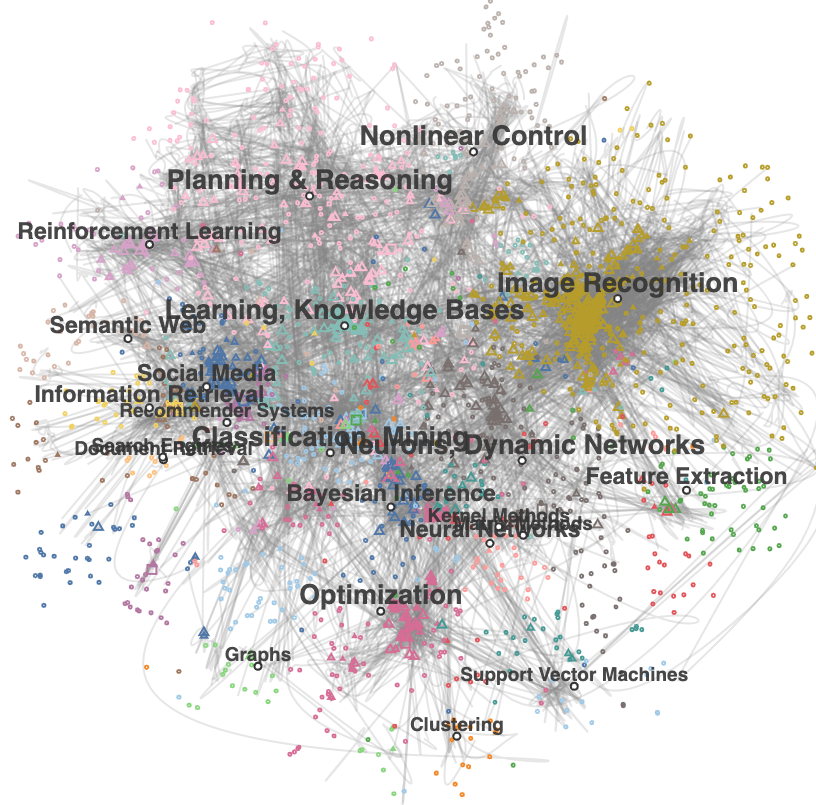}
\caption{Trajectory Overview. In this view, we display all trajectories from our data set together instead of only one trajectory instance, revealing more of the overall trend of where trajectories are located. We did not include this view in the final demo application due to performance reasons and the reduced overview.}\label{fig:trajectoryoverview}
\end{figure}

\begin{figure}
\includegraphics[height=0.5\columnwidth,trim={2cm 0cm 1cm 0cm},clip]{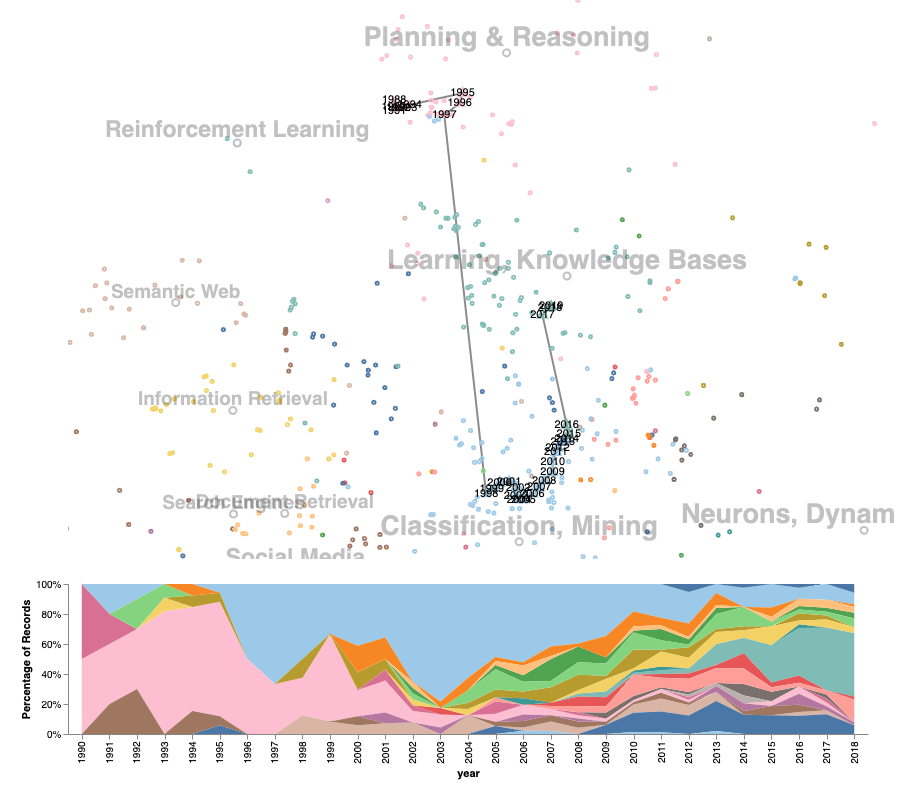}
\includegraphics[height=0.5\columnwidth,trim={2cm 0cm 2cm 0cm},clip]{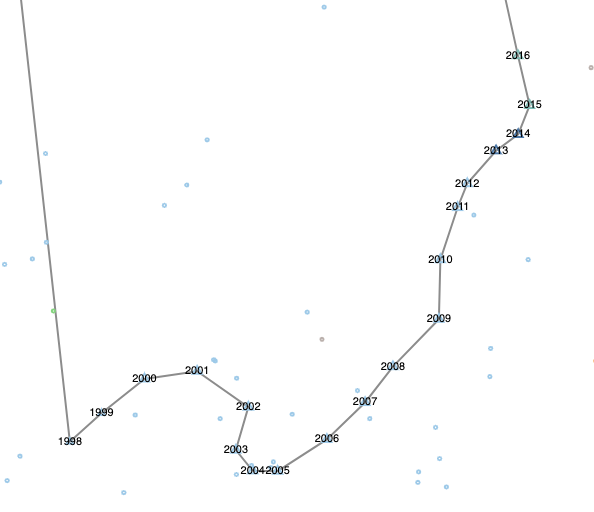}
\caption{Overview (left) and closeup (right) of an author trajectory (\emph{Jiawei Han}). In this view, spline interpolation is turned off to show that there is some smoothness in the trajectory already without it. In the map overview to the left, one may notice three main stations in the author's career (\emph{Planning \& Reasoning}, \emph{Classification, Mining} and \emph{Knowledge Bases}). This is confirmed and complemented by the stream graph below the map. The closeup to the right shows the part of the trajectory located in the \emph{Classification, Mining} area. Here, different parts of the area may also be related to different subtopics.}\label{fig:jiaweihancloseup}
\end{figure}
\begin{figure}
\centering
\includegraphics[height=0.56\columnwidth,trim={0.5cm 0cm 0cm 0cm},clip]{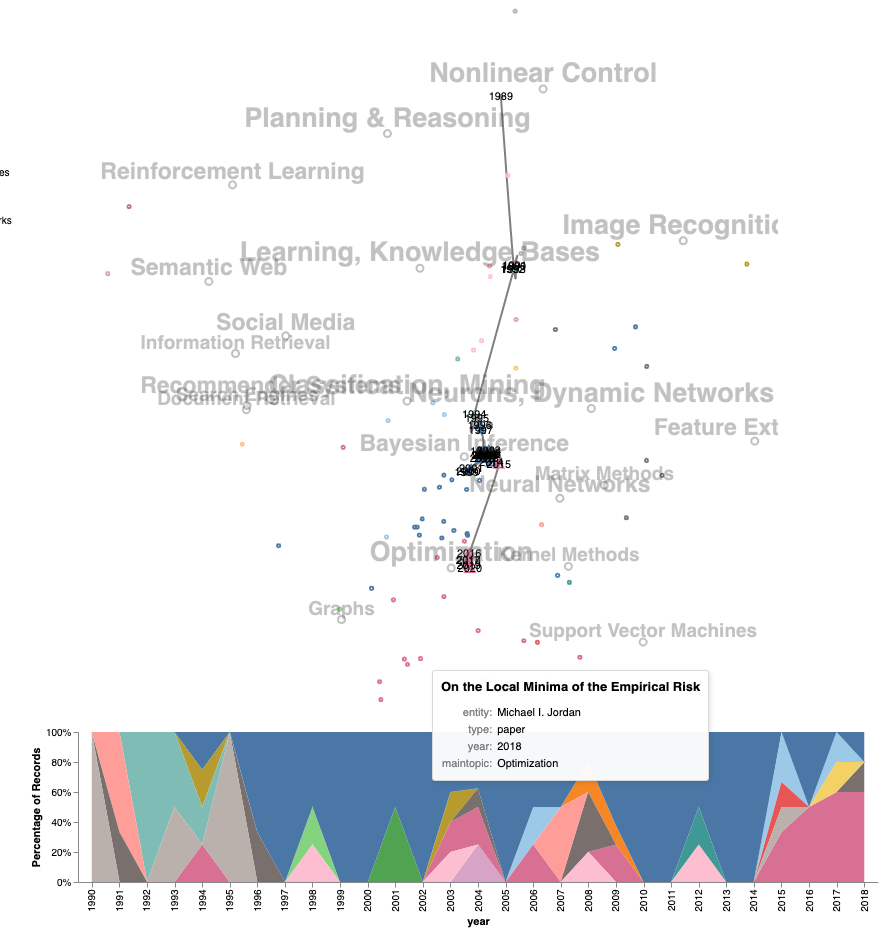}
\includegraphics[height=0.56\columnwidth,trim={0cm 0cm 0cm 0cm},clip]{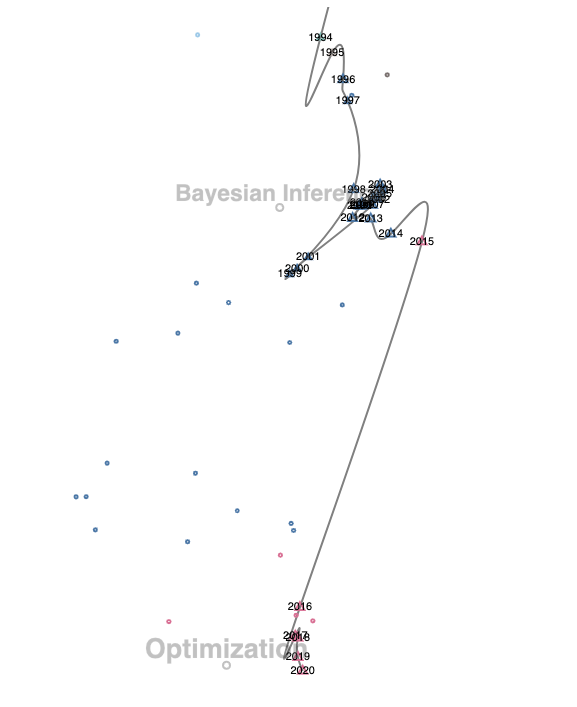}
\caption{Overview (left) and closeup (right) of an author trajectory (\emph{Michael Jordan}). Spline interpolation is turned on. The author started in \emph{Nonlinear Control}, moving to \emph{Knowledge Bases}, \emph{Bayesian Inference}, and, finally, \emph{Optimization}. An exemplary tooltip of a paper nearby the optimization part of the trajectory confirms the discovered topic.}\label{fig:michaeljordancloseup}
\end{figure}

\end{document}